\documentclass[aps,prl,preprint]{revtex4}
\usepackage{graphicx}
\usepackage{bm}
\usepackage{subfigure}
\usepackage{amsmath,amsfonts,amssymb}
\DeclareGraphicsExtensions{.eps}
\graphicspath{{c://share/users/IBM/public/research/droplet_collision/dropcollision2/analysis/fig/}{c://share/users/IBM/public/research/droplet_collision/dropcollision2/logs/}}

\begin{document}
%%%%%%%%%%%%%%%%%%%%%%%%%%%%%%%%%%%%%%%%%%%%%%%%%%%%%%%%%%%%%%%%%%%%%%%%%%%%
\title{Second Blows in the Head-on Collisions of the Spherical Nano Polymer Droplets}
%%%%%%%%%%%%%%%%%%%%%%%%%%%%%%%%%%%%%%%%%%%%%%%%%%%%%%%%%%%%%%%%%%%%%%%%%%%%
%
\author{Sangrak Kim}
\email[E-mail address: ]{srkim@kgu.ac.kr}
\affiliation{Department of Physics,\\ Kyonggi University,\\ 94-6 Eui-dong, Youngtong-ku, Suwon 440-760, Korea}
\date{\today}
%\date{Received: date / Revised version: date}
%
%%%%%%%%%%%%%%%%
\begin{abstract}
%%%%%%%%%%%%%%%%

We report observations of weird but interesting phenomenon from the molecular dynamics simulations, occurrence of second blows in the head-on collisions of two equal-sized spherical nano polymer droplets. In the head-on collisions, we usually expect a single peak of impact forces between two colliding droplets. But, in the simulations, the second peak of the impact forces is actually observed. Its underlying mechanisms at the molecular scale are also proposed.

%%%%%%%%%%%%%%%%
\end{abstract}
%%%%%%%%%%%%%%%%


\maketitle
%
%%%%%%%%%%%%%%%%%%%%%%%
\section{introduction}
\label{sec:introduction}
%%%%%%%%%%%%%%%%%%%%%%%
Droplets collisions are a very common phenomenon in nature or in industry applications as well. For example, the changes in water droplets sizes by their collisions result in quite different weathers ranging from heavy rain falls to thin fogs \cite{rain}. Furthermore, its understanding enables us to better control the droplet processing in applications, such as spray processing \cite{spray1}-\cite{spray2}, microfluidics \cite{microfluidics1}-\cite{microfluidics2}, and so on.

Many investigations of the droplet collision dynamics have been done by using experimental, simulation, and theoretical approaches \cite{sun}-\cite{inamu}. For the case of experiments of nano droplets collision, it is very hard to precisely control the experimental setups and/or parameters, such as their sizes, shapes, impact velocities, and impact parameters, \textit{etc}. Understanding of droplets collision, in particular, at a molecular level, is still very poor. Simulation has some advantages over the experiment and theory, and can easily and precisely control experimental setups and/or parameters. It can thus significantly enhance the detailed understanding of the collision dynamics inside the droplets which cannot be readily accessible in experiments. In this study of nano droplets collision, therefore, we employ a simulation approach \cite{md1}-\cite{prev}.

Here, we report an interesting behavior of molecular dynamics (MD) simulation results, which is related to the impact forces between two colliding droplets. Intuitively, when two droplets collide head-on, they experience impact forces from the other and show a single peak of the impact forces during the collision. However, in our molecular dynamics simulations, unexpected occurrences of a second peak in the impact force are observed. This is called a phenomenon of second blow. As the impact speed is lowered, the two peaks converge into one and finally disappears. This second blow has not yet been reported elsewhere.

This article is organized as follows: In section~\ref{sec:method}, we first describe the simulation method and relevant simulation parameters. In section~\ref{sec:results}, simulation results on the second blows are reported and the relevant parameters to characterize them are introduced. Finally, discussions and concluding remarks are followed.

%%%%%%%%%%%%%%%%%%%%%%%
\section{simulation method}
\label{sec:method}
%%%%%%%%%%%%%%%%%%%%%%%
Let us first describe the simulation model \cite{prev}. The polymer chains in the droplets are modeled by a bead-spring model \cite{sides}. The chain length of a polymer is $L ~=~ 10$. All monomers in the same droplet interact with each other through the Lennard-Jones (LJ) potential, given as
\begin{equation}
    V_{LJ}(r) ~=~ 4\epsilon[(r/\sigma)^{-12}- (r/\sigma)^{-6}]. \label{ljpot}
\end{equation}
All the quantities are expressed in Lennard-Jones reduced unit, just for convenience. Neighboring monomers in the same chain additionally interact with the finite extension non-linear elastic (FENE) potential given by,
\begin{equation}
    V_{F}(r) ~=~ -\frac{k_F}{2} r_0 ~\ln[1-(r/r_0)^{2}], \label{fene}
\end{equation}
where $k_F$ is a spring constant and $r_0$ is a maximum length within which the chain can be sustained. In our simulations, we choose $r_0 ~=~ 1.5$ and $k_F ~=~ 30.0$. The interaction between molecules in the different droplets is set to be completely repulsive.

We prepare two colliding spherical droplets as follows: First, we make a polymer droplet in a state of melt such that the positions and velocities correspond to a thermodynamic state with a density $\rho ~=~ 0.5$ and a temperature $T ~=~ 0.5$. The shape of the droplet is initially cubic. To make a spherical droplet from cubic one, we connect each molecule with a virtual harmonic spring with a weak spring constant and the other end of each spring is connected to the center of the simulation box to contract the droplet towards its center. We can eventually have a spherical droplet with a radius $R ~=~ 22.87$ and a thermodynamic state of a density $\rho ~=~ 0.83$ and a temperature $T ~=~ 0.5$. After the shape of droplet becomes spherical, we disconnect the harmonic spring and then equilibrate for some time. In order for the droplets to collide, they are assigned the same approach speed $u$, so their relative speed corresponds to $2 u$ along the x-axis. We choose a significantly larger value of the cutoff length $r_c ~=~ 5.0$ for the LJ interaction, compared to the usual value of  $r_c ~=~ 2.5$, in order to avoid an unwanted force, when they enter the interaction cutoff region from a distance. The number of molecules composing a droplet is $N ~=~ 40,000$. We average over 32 different initial configurations by rotating each droplets.

%%%%%%%%%%%%%%%%%%%%%%%
\section{simulation results}
\label{sec:results}
%%%%%%%%%%%%%%%%%%%%%%%
\begin{figure}[htbp]
\begin{center}
\includegraphics[scale=0.6]{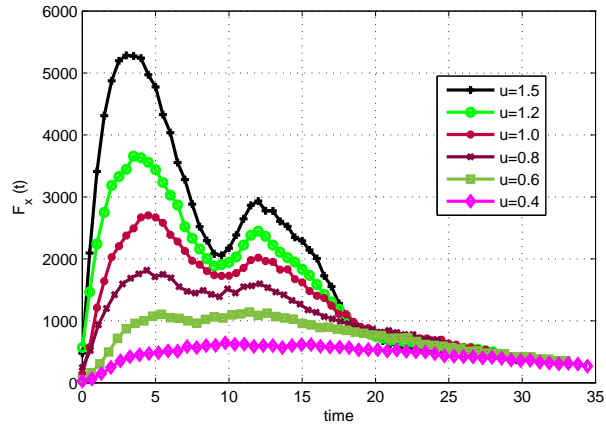}
\caption{(Color on-line) Impact force $F_x (t)$ for different impact speeds $u$. We can clearly see the second blows when $u > 1.0$.}
\label{mrfx}
\end{center}
\end{figure}

\begin{table}
\caption{Height ratio $H$ and parameter $S$ with the impact speed $u$. As the impact speed $u$ is larger, the second blow becomes more pronounced. The second blows are well classified if $H>2.5$ and $S>2.0$.}
 \begin{tabular}{||c|c|c||} \hline
 u & H & S \\\hline
 1.5 & 3.69 & 4.11 \\\hline
 1.2 & 3.14 & 3.70 \\\hline
 1.0 & 3.32 & 2.46 \\\hline
 0.8 & 2.05 & 1.85 \\\hline
 0.6 & 0.51 & 2.50 \\\hline
 \end{tabular}
\label{distances}
\end{table}

When two droplets collide, the total impact forces $F_x (t)$ between them are exerted each other. These forces are averaged over 32 different initial configurations. The time $t = 0$ is set when two colliding droplets just begin to feel the impact forces $F_x$. The simulation results are shown in Fig. \ref{mrfx} for different impact speeds $u$.

Let us consider the case $u = 1.5$, as a specific example. The first peak in impact forces occurs at $t=3.0$ with a total force $F_x =5283$ and the second peak at $t=12.0$ with $F_x =2933$. The local minimum between the first and the second peak occurs at $t=9.5$ with $F_x =2058$. In this case, we can clearly see the second blow.

As the impact speed decreases, the times at which the first peak and second peak occur come closer each other and eventually converge into one. The relative magnitude of the first peak and the second peak are also reduced. At impact speed $u = 0.4$, there is only one peak in $F_x (t)$.

To characterize the second blow phenomenon, we introduce two parameters, $H$ and $S$. The relative height between the first peak and the second peak with respect to the local minimum of the impact force is defined as relative force magnitude ratio
\begin{equation}
    H \equiv \frac{F_1 - F_v}{F_2 - F_v}, \label{H}
\end{equation}
where $F_1$ is the first peak force, $F_2$ is the second peak force, and $F_v$  is the local minimum force. The second blow can also be characterized by a quantity $S$, which is defined as
\begin{equation}
    S ~=~ \frac{F_2 - F_v}{s}, \label{S}
\end{equation}
where $s$ is an average of standard deviations of the impact forces $F_x$ over 32 different initial configurations. Thus, $S$ can be used to find how conspicuous the second peak is relative to the local minimum. The calculated results are summarized in Table \ref{distances}.

As the impact speed $u$ is lowered, the second peak gradually disappears. Below impact speed $u = 0.8$, it is rather difficult to identify the second peak, having $H < 2.5$ and $S < 2.0$. On the contrary, at impact speed $u=1.5$, we have $H = 3.7$ and $S = 4.1$.

%%%%%%%%%%%%%%%%%%%%%%%
\section{discussions and conclusions}
%%%%%%%%%%%%%%%%%%%%%%%
As the two droplets begin to collide, the impact forces $F_x$ between two droplets are rather rapidly increased up to the first peak and then it is decreased for a moment, and then again it is slowly increased to make the second peak. When the impact speed is small, the impact forces do not show any second peak. But, when the impact speed is greater than $u = 0.8$, they begin to show the second peak. The second peak is always followed by the first peak. The third peak is not observed in the simulations. As shown in Fig. \ref{mrfx}, the interval between the two peaks is wider as impact speed increases. The times at which the first peak occurs is smaller and the height of the first peaks is larger as the impact speed increases.

Let us now consider the underlying mechanism of the second blow phenomenon. This second blow seems to be due to two mechanisms. The droplets are in melt state, rather than in rigid solid state. Even after the first blow occurs, the area of contact area increases. Thus, more molecules involve in the collision process. This supplies newly incoming molecules in the outer region. However, the momenta and kinetic energy supplies are significantly reduced in the region near the collision axis. Furthermore, new incoming momenta are mainly supplied from the molecules around at $r=\pm15.0$ from the colliding center of the droplets. The combination of participating number and new kinetic energy supplies make the second blow occur around at $t=12.0$. This makes a spill-over effect at the time of the second blow. From these facts, we can also infer that we cannot observe any second blow, if the radius $R$ of a droplet is less than 10.0.

In the simulations, we fix the polymer chain length unchanged even under strong collision. However, in reality, when two polymer droplets collide strongly, the polymer chain may be broken into parts. The polymer chain transfers the kinetic energy of the droplets into the elastic potential energy. The polymer chain saves elastic potential energy. This elastic potential energy reduces the impact forces between two colliding droplets, showing a local minimum of the impact forces. Later, when the elastic potential energy is restored into kinetic energy, consequently increase the impact force between droplets. This is a kind of slingshot effect, related with the elastic behavior of the polymer chain. This may be an artefact of the simulations, because the polymer chain is kept unbroken as they collide in the simulations. This implies that if polymer chain breaks in the strong collision, this phenomenon may not happen. Actually, the impact speed is larger than $u=1.6$, there are some chances to break droplets into parts for some initial configurations. The time for the molecules at the surface of the droplet to move freely along the impact direction is around at $t = 7.5$. This may give the lower bound of the time for the second blow to occur.

In summary, we present the second blow phenomenon during the head-on collisions of spherical nano polymer droplets through the molecular dynamics simulations and proposed the underlying mechanisms of this phenomenon at the molecular scale. This second blows are observed for different initial configurations. This is checked over 32 different initial conditions. The possible mechanisms for second blow may be due to the spill-over effect and/or the slingshot effect. The phenomenon may be explained by a combination of these two mechanisms. Finally, we raise the problem and further investigations on the subject should be followed.

\end{document}